# Towards a Heavy-ion Transport Capability in the MARS15 Code[*]

N.V. Mokhov[1], K.K. Gudima[2], S.G. Mashnik[3], I.L. Rakhno[4], S.I. Striganov[1]

[1]*Fermi National Accelerator Laboratory, MS 220, Batavia, Illinois 60510-0500, USA*
[2]*Institute of Applied Physics, Academy of Sciences of Moldova, Kishinev, MD-2028, Moldova*
[3]*Los-Alamos National Laboratory, MS B283, Los-Alamos, New-Mexico 87545, USA*
[4]*University of Illinois at Urbana-Champaign, 1110 W. Green St., Urbana, Illinois 61801-3080, USA*

April 29, 2004

## Abstract

In order to meet the challenges of new accelerator and space projects and further improve modelling of radiation effects in microscopic objects, heavy-ion interaction and transport physics have been recently incorporated into the MARS15 Monte Carlo code. A brief description of new modules is given in comparison with experimental data.

# TOWARDS A HEAVY-ION TRANSPORT CAPABILITY IN THE MARS15 CODE[*]

N.V. Mokhov[1][**], K.K. Gudima[2], S.G. Mashnik[3], I.L. Rakhno[4], S.I. Striganov[1]

[1]*Fermi National Accelerator Laboratory, MS 220, Batavia, Illinois 60510-0500, USA*
[2]*Institute of Applied Physics, Academy of Sciences of Moldova, Kishinev, MD-2028, Moldova*
[3]*Los-Alamos National Laboratory, MS B283, Los-Alamos, New-Mexico 87545, USA*
[4]*University of Illinois at Urbana-Champaign, 1110 W. Green Street, Urbana, Illinois 61801-3080, USA*

**Abstract.** In order to meet the challenges of new accelerator and space projects and further improve modelling of radiation effects in microscopic objects, heavy-ion interaction and transport physics have been recently incorporated into the MARS15 Monte Carlo code. A brief description of new modules is given in comparison with experimental data.

## INTRODUCTION

The MARS Monte Carlo code[1] is widely used in numerous accelerator, detector, shielding and cosmic ray applications. The needs of the Relativistic Hevy-Ion Collider, Large Hadron Collider, Rare Isotope Accelerator and NASA projects have recently induced adding heavy-ion interaction and transport physics to the MARS15 code. The key modules of the new implementation are described below along with their comparisons to experimental data.

## HEAVY-ION NUCLEAR INTERACTIONS

The 2003 version[2] of the LAQGSM code[3] was implemented into MARS15 after substantial revisions and merging with CEM03 and native MARS14 modules. It can now be used in full transport simulations in complex macro-systems for modelling all heavy-ion and hadron nuclear interactions from 10 MeV/A to 800 GeV/A as well as photo-nuclear interactions (currently in a limited energy range). LAQGSM03 includes an improved version of the Dubna intra-nuclear cascade model[4] that makes use of experimental elementary cross sections (or those calculated with the Quark-Gluon String Model for energies above 4.5 GeV/A), the pre-equilibrium model from the improved Cascade-Exciton Model[5] as realized in the CEM03 code[6,7], refined versions of the Fermi break-up and coalescence models[4], and an improved version of the Furihata's Generalized Evaporation-fission Model[8] (GEM2). This implementation provides a power of full theoretically consistent modelling of exclusive and inclusive distributions of secondary particles, spallation, fission, and fragmentation products. Although benchmarking results, shown in Figs. 1 and 2 for nuclide production and in Figs. 3 and 4 for inclusive spectra, are quite impressive, further development of this package is underway[6,7].

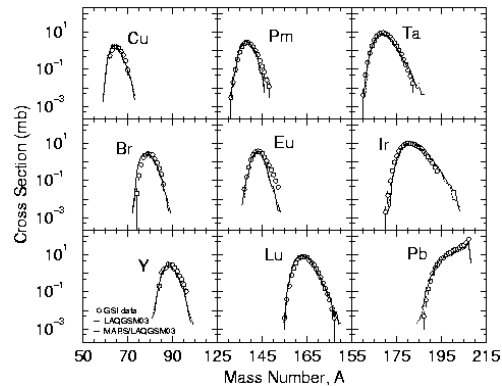

Figure 1. Mass yield in $d+^{208}Pb$ interaction at 1 GeV/A as calculated with original LAQGSM03 and that implemented into MARS15, and measured in Ref. (9).

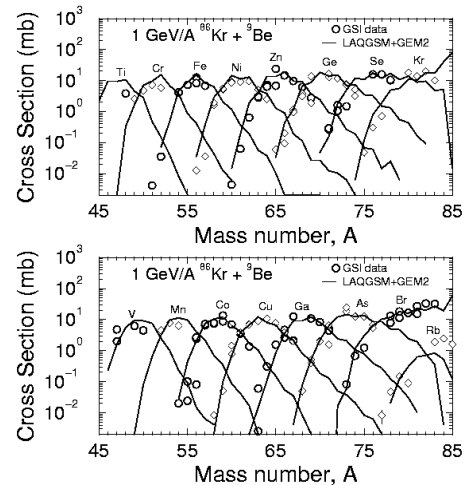

Figure 2. Mass yield in $^{86}Kr + ^{9}Be$ reaction at 1 GeV/A as calculated with LAQGSM and measured in Ref. (10).

[*] This work was supported by the Universities Research Association, Inc., under contract DE-AC02-76CH03000 with the U.S. Department of Energy, and in part by the Moldovan-U.S. Bilateral Grants Program, CRDF Projects MP2-3025 and MP2-3045, the NASA ATP01 Grant NRA-01-01-ATP-066, and the Higher Education Cooperative Act Grant of the Illinois Board of Higher Education.
[**] mokhov@fnal.gov

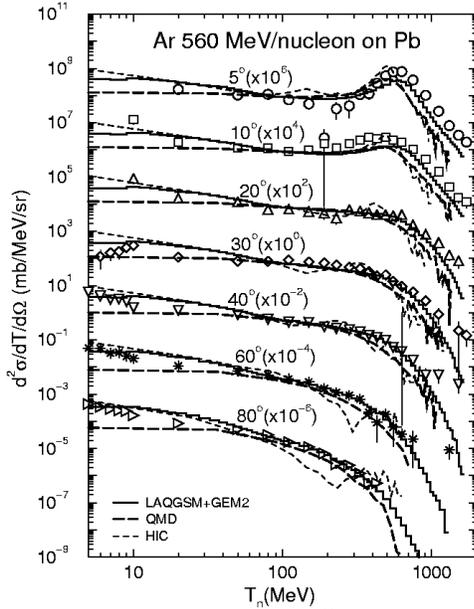

Figure 3. Comparison of measured[11] differential cross sections of neutrons with LAQGSM and calculations with QMD and HIC models for 560 MeV/A *Ar+Pb* reaction.

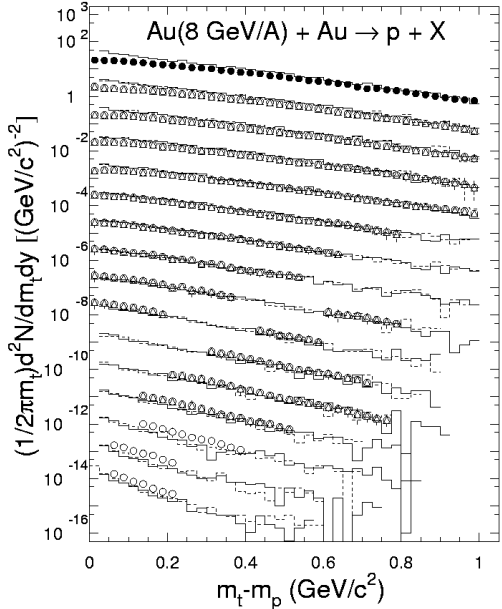

Figure 4. Invariant proton yield per central *Au+Au* collision at 8 GeV/A as calculated with LAQGSM03 (histograms) and measured in Ref. (12) (symbols). Solid lines and open circles is forward production, dashed lines and open triangles is backward production. Midrapidity (upper set) is shown unscaled, while the 0.1 unit rapidity slices are scaled down by successive factors of 10.

## NUCLEUS-NUCLEUS CROSS SECTIONS

Inelastic and elastic cross sections for heavy ion nuclear interactions have been extensively studied both theoretically and experimentally for the past decades. Nevertheless, quality of experimental data is still far from perfect and measurements by different groups in many cases contradict each other[13]. Several empirical prescriptions have been developed for inelastic cross sections. Most recent and comprehensive studies were performed at NASA[14] and JINR[15]. Both models show reasonable agreement with experimental data for energy range from a few A MeV to 200 A GeV and for ions, both projectiles and targets, ranging from deuteron to lead. A typical comparison between data and these models is presented in Fig. 5 for $^{20}$Ne ions. Situation is similar for other projectiles. The JINR model was chosen for MARS15 because it describes also elastic cross sections needed for full particle transport.

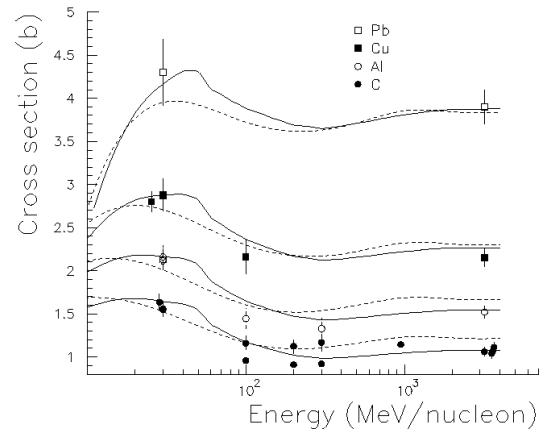

Figure 5. Data[13] (symbols) on inelastic cross sections of $^{20}$Ne ions *vs* JINR[15] (solid) and NASA[14] (dashed) models.

## ELECTROMAGNETIC INTERACTIONS

A latest version of the algorithm used in MARS15 for modelling of correlated ionization energy loss and multiple Coulomb scattering is described in Ref. (16). It is based on a separate treatment of "soft" and "hard" interactions. A large number of "soft" collisions are described using a "continuous scattering approximation"; a small number of "hard" collisions are simulated directly. A transition angle between "soft" and "hard" collisions is determined as a function of a step-size providing a possibility for fast and precise simulation. "Hard" collisions are sampled taking into account projectile and target form-factors and exact kinematics of projectile-electron interactions. Corresponding energy losses are calculated from a simulated scattering angle. Energy loss in "soft" projectile-electron collisions is described by a modified Vavilov function, approximated by a log-normal distribution. Calculated correlations between energy loss and scattering are quite substantial for low-Z targets. The mean stopping power used for a normalization of ionization energy loss on a step is described in the next section. This algorithm is directly applicable for heavy ions of a charge $z$. Measurements on various targets[17] confirm this, although there is an evidence that the width of energy loss distributions is narrower for $z$ close to 100. Radiative processes for heavy ions – bremsstrahlung and direct pair production – are modelled directly[1].

## MEAN STOPPING POWER

The mean ionization energy loss, $dE/dx$, is calculated using the Bethe-Bloch formalism[18] in combination with various corrections[1,18]. Two additional corrections have been implemented into the code to describe better the Barkas effect and take into account electron capture by low-energy ions. An improved Barkas term is described according to Ref. (19). Electron capture reveals itself as a significant decrease in an ion charge at low kinetic energies. The effective ion charge, $z_{eff}$, is determined according to semi-empirical formulae[20] and used instead of a bare projectile charge. Thus, the regular ionization logarithm[18]

$$0.5\ln(2m_e c^2 \beta^2 \gamma^2 T_{max} I^{-2}) - \beta^2$$

is multiplied by the factor of $1+z_{eff}F(V)/\sqrt{Z}$, where $F(V)$ is a tabulated function, $V$ is a dimensionless parameter equal to $\beta\gamma/\alpha\sqrt{Z}$, $\alpha$ is the fine-structure constant, and all the other terms have their usual meaning[18-20]. A comparison with data is shown in Figs. 6-7.

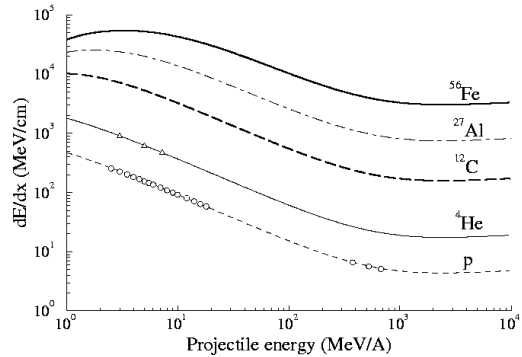

Figure 6. Measured[21] (symbols) and calculated (lines) ionization energy losses by various ions in aluminium.

For light ions like $^{12}$C our model is well justified – in aluminium the difference in the calculated $dE/dx$ values between our approach and a more comprehensive (from formal standpoint) one[19] is less than 1% in the energy region from $10^{-3}$ up to $10^5$ GeV/A. For Pb ions the difference increases up to 6% above 0.1 GeV/A. The advantage of our approach is in its speed (a factor of $10^3$ when compared to Ref. (19)).

Bremsstrahlung and direct pair production by heavy ions are modelled directly, as mentioned in previous section, or – alternatively – can be treated in a continuous slowdown approximation with corresponding MARS modules for $dE/dx$ (see Ref. (1,18)).

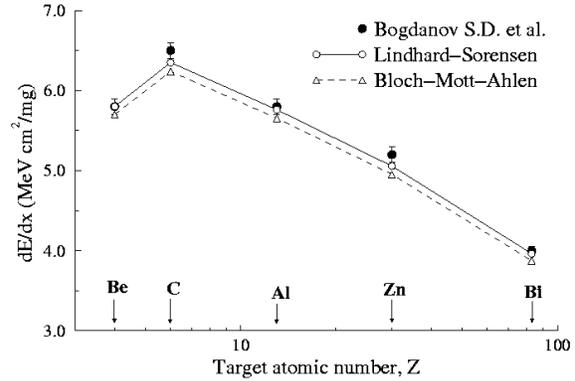

Figure 7. Measured[22] and calculated (full and open symbols, respectively) ionization losses by 780-MeV/A Xe ions.

## MARS15 TEST FOR A THICK LEAD TARGET

Results of a heavy-ion implementation test in MARS15 are presented in Fig.8 for a lead cylinder of 20-cm diameter and 60-cm thick irradiated by 0.5 to 3.65 GeV/A light ion beams. Our results are in a good agreement with data[23] and predictions of the latest version of the SHIELD code[24] presented in Ref. (25).

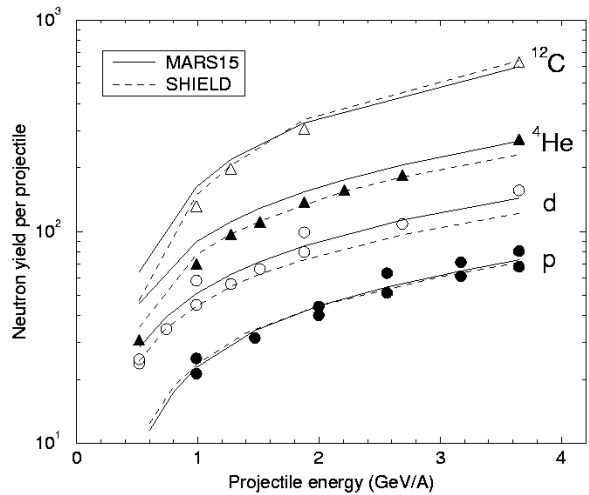

Figure 8. Calculated and measured[23] total neutron yield (E<14.5 MeV) from a lead cylinder *vs* ion beam energy.